\documentstyle[multicol,aps,prl,epsf]{revtex}
\begin{document}
\title{Structure and Magnetism of Neutral and 
Anionic Palladium Clusters }
\author{  M. Moseler, 
                H. H\"akkinen, 
                R.N. Barnett,
            and Uzi Landman,\\
School of Physics, Georgia Institute of Technology\\
Atlanta, GA 30332-0430}
\date{September 28, 2000}
\maketitle
%
%
\begin{abstract}
The properties of neutral and anionic Pd$_N$ clusters were investigated
with spin-density-functional calculations. The ground state structures 
are three-dimensional for $N$$>$3 and they are magnetic with a spin-triplet
for 2$\le$$N$$\le$7 and a spin nonet for $N$$=$13 neutral clusters.
Structural- and spin-isomers were determined and an anomalous increase 
of the magnetic moment with temperature is predicted for a Pd$_7$ ensemble.
Vertical electron detachment and ionization energies were calculated
and the former agree well with measured values for Pd$_N^-$.
\pacs{PACS: 36.40.Cg, 36.40.Mr, 36.40.Wa, 61.46.+w}
\end{abstract} 
\begin{multicols}{2}
%
%
%
%
%

Enhancement of magnetism 
in finite clusters made of elements that are ferromagnetic as bulk
solids has been demonstrated  through Stern-Gerlach (SG)
deflection measurements~\cite{Billas},
and it is understood to derive from reduced atomic 
coordination resulting in stronger electron localization.
However, the emergence of 
magnetism in small clusters  of close-shell non-magnetic atoms is
clouded by  uncertainty. 
Palladium aggregates (with a  [Kr] 4d$^{10}$
structure of the atom) are particularly 
interesting  since the bulk metal is known to be "almost" magnetic, with 
the emergence of magnetism predicted to require a mere 6 \% dilation of 
the interatomic distance~\cite{Moruzzi}. 
Experimental information  is 
limited to SG measurements on Pd$_N$ clusters
with $N$$>$12 (where the clusters were found to be non-magnetic)~\cite{Cox} or 
to inferences from (non-magnetic) photoemission studies~\cite{Gantefoer}.
Additionally,  systematic theoretical investigations of magnetic properties 
are lacking, limited only to neutral
 Pd$_{13}$~\cite{Reddy,Watari}.

We report on an extensive density-functional-theory (DFT) study
pertaining to  size-dependent evolutionary patterns of the 
 properties of Pd$_N$ and
Pd$^-_N$ clusters with $N$$=$1$-$7 and $N$$=$13.
 The ground states (GSs) of the neutral and anionic clusters 
are found to undergo  an early transition (i.e. for $N$$>$3)
to three-dimensional (3D) ones and  possess a non-zero magnetic moment
(e.g., for neutral clusters with $N$$\le$7 a spin triplet, $S$$=$1,  
and a spin nonet, $S$$=$4,
for $N$$=$13). In addition to higher-energy structural isomers
(STIs) we determined sequences of close-lying spin isomers (SPIs).
For a Pd$_7$ ensemble, we predict an increase of the magnetic
moment with temperature due to the thermal accessibility of such isomers.
All the SPIs exhibit  high local magnetic moments (LMMs), 
including the singlet states with an
 antiferromagnetic  LMM coupling.
Our structural determinations are corroborated
by the remarkable agreement between the calculated vertical electron
detachment energies (vDEs) from the cluster anions and
photoelectron spectroscopy (PES) 
measurements\cite{Gantefoer,Lineberger1,Lineberger2}.

In this study the Kohn-Sham (KS) equations with
generalized gradient corrections (GGA) \cite{Perdew}
were solved using the Born-Oppenheimer local-spin-density
molecular dynamics (BO-LSD-MD) method~\cite{Barnett},  with
scalar-relativistic~\cite{Kleinman}
 non-local  pseudopotentials~\cite{Troullier}. 
Cluster geometries were determined  via symmetry-unrestricted
structural optimizations through a  
rather exhaustive  search
starting from various structures including those 
suggested in previous studies
of certain Pd clusters \cite{Balasubramanian,Valerio,Neyman}
 and other metal
clusters\cite{Hakkinen,Bonacic-Koutecky}.
For each structure
spin-restricted optimizations were performed 
covering all energetically
important spin multiplicities.

The GS geometries and STIs for both Pd$_N$ and Pd$^-_N$ are rather
similar and they follow  the same structural evolution, exhibiting
Jahn-Teller distortions from the ideal symmetric
 structures (see Fig. 1), with a transition
to 3D configurations at $N$$>$3~\cite{Note1}.
The  GSs (i.e. maximal binding energy $E_B$ in Fig.~1)
 of Pd$_N$ with $N$$\le$7
have a triplet spin multiplicity (see open squares
in Fig. 1(A-F)), while for
Pd$_{13}$ the GS is associated with a nonet ($S$$=$4) spin configuration
(see Fig. 1G). On the other hand, the GS spin-multiplicities of the
Pd$^-_N$ anions (filled symbols in Fig. 1) vary non-monotonically
with $N$ (doublet for $N$$=$2,3,4, and 6; quartet for $N$$=$5;
sextet for $N$$=$7, and octet for $N$$=$13).
We note here that the higher-lying STIs\cite{Note2} and SPIs of the neutral
and anionic clusters become thermally accessible with increasing
cluster size (see temperature scales \cite{Temp} on the right-hand-side
in Fig. 1).

The binding energies of the GS
clusters increase rather monotonically with $N$
(Fig. 2a), showing enhanced local stabilities 
for Pd$^-_3$ and Pd$_4$~\cite{Note3}. The average nearest-neighbor
bond-length ($\langle R_{nn}\rangle$)
 of the GS structures  and the SPIs
are very similar \cite{Note4}, converging rapidly to the bulk value 
(2.75 \AA~\cite{Kittel}).

The calculated vDEs from the cluster anions are in a remarkable
agreement with values  
determined from PES measurements \cite{Gantefoer,Lineberger1,Lineberger2}
 (Fig. 2b). For all the
clusters, we display the vDEs only for  the GS (structural and
spin) configurations of the anions, 
except for Pd$^-_7$ where we show the vDEs
for both the GS ($S=5/2$) and its 
next (higher-in-energy) SPI ($S=3/2$) (the difference in energy
between the two isomers is 0.11 eV, corresponding to only
170 K in vibrational temperature \cite{Temp}), with the latter exhibiting
a better agreement with the experimental data.
These results suggest that for Pd$^-_7$ the measured vDE  may
correspond to this higher-energy SPI which is accessible already
at very low temperatures, while for the other clusters
studied here the vDEs  are determined by the GS structures 
over a broad temperature range. 

The only (slight) discrepancy between the calculated and measured
VDEs is for  Pd$^-_3$ (see Fig. 2b). However,
 we observe from this figure that the predicted  vDE (2.0~eV)
for Pd$^-_3$ is rather close to the dissociation  energy ($E_D$) 
of the process Pd$^-_3$$\rightarrow$Pd$^-_2$$+$Pd
(see filled circle in Fig. 2b at $E_D$$=$2.41~eV, compared to
a measured value of 2.26~eV~\cite{Spasov}).
This implies that this dissociation channel may compete
with the electron detachment under appropriate experimental conditions.
Consequently, we 
assign the vDE of Pd$_3^-$ to the next higher feature 
in the measured PES~\cite{Lineberger2}) occuring at 1.88~eV which
agrees with the predicted value, 
 and the lower measured value
(at 1.66~eV, marked by a cross in Fig. 2b)
is attributed  to the vDE of a (hot) Pd$^-_2$ 
dissociation product \cite{Pd2} 
(the same conclusions apply to the photoelectron spectrum of Pd$^-_3$
shown in Fig. 2 of ref.~\cite{Gantefoer}); 
for a similar interpretation 
of  the measured PES of Au$^-_3$ see Refs. 16 and 26.

The vertical ionization potentials  predicted by
us (vIP in Fig. 2c) start at the atomic
value of 8.28 eV compared to the measured first
IP of Pd, 8.33 eV~\cite{Kittel}; the calculated second IP of Pd (27.34~eV)
also agrees with the experimental one (27.75~eV)~\cite{Kittel}.
The predicted
values for $N$$=$2 and 3 are close to each other followed by a marked drop
for clusters with $N$$>$3, which is likely to reflect the
transition to 3D structures. Convergence to the bulk limit (the
work function of Pd is 4.97~eV \cite{Handbook}) is  slow.
No measured IP values have been reported.

The multitude of   spin-multiplicities for the Pd$^-_N$
clusters results in  a considerable 
variation of the 
magnetic moment per atom $\mu$$=$$2S\mu_B/N$ 
(solid squares in Fig.~2d) with particularly 
 high values of $\mu$$=$0.6, 0.71 and 0.54 $\mu_B$  
for $N$$=$5, 7 and 13, respectively. 
On the other hand, the occurrence
of a  triplet GS for the neutral clusters underlies
  a monotonic $1/N$ 
decrease of $\mu$  for $N$$=$1$-$7 
(open squares in Fig.~2d), with  an unexpected high value of 0.62~$\mu_B$
for  Pd$_{13}$ which is higher than the experimental~\cite{Cox}
estimate ($\mu$$<$0.4~$\mu_B$).

The rather surprising existence of energetically favorable high-spin
isomers for Pd clusters originates from $sd$-hybridization
developing upon bonding, with the $d$-orbitals' weight
diminished somewhat on each atom, resulting in a situation
reminiscent of open-shell transition metals.
 The total magnetic moment of the cluster is comprised of 
sizable atomic LMMs ($\mu_\ell$$\sim$$\pm(0.3 - 0.6) \mu_B$)
that couple  antiferromagnetically  
in the spin-compensated singlet states, and align 
 themselves in 
 SPIs with high total $\mu$. 
This is illustrated for the icosahedral Pd$_{13}$
cluster in Fig. 2 where the (gapless) density of states
(DOS) of the $S=0$ singlet SPI is shown in 
Fig. 2e and the corresponding spin-polarization density
in Fig. 2g (left); note the non-uniform spatial
distribution of the spin-polarization reflected in the
different line-shapes of the DOS of the up and down
spins. In the $S=2$ spin-quintet SPI the minority
spin polarization is localized on three sites located
in a triangle (Fig. 2g, middle). In the $S=4$
spin-nonet GS cluster all the sites are spin-polarized
in the same direction (Fig. 2g, right), and it's stability 
is reflected in the large gap
in the majority-spin DOS near the Fermi energy
(see spin $\uparrow$ in Fig. 2f). 

Intriguing conclusions can be made regarding the thermal behavior
of certain Pd clusters in SG measurements;
here we consider the case of Pd$_7$ and
Pd$^-_7$ at room temperature.
 Neglecting 
the vibrational and entropic differences   
between the isomers, the probability to find a cluster 
with spin $S$, irrespective of its atomic isomeric structure, 
is given by 
$P_{N,T}(S)=\sum_I \exp{(\frac{N E_B(N,I,S)}{k_B T})}/Z_{N,T}$
with the Boltzmann constant $k_B$, the ensemble 
temperature $T$,
the structural isomer index $I$,  and the (normalizing) 
partition function $Z_{N,T}$. Several spin isomers 
of the neutral as well as the anionic heptamer 
have a finite $P_{7,T}(S)$ (Fig.~3a) even for 
low temperatures,  leading us to predict that in a SG
experiment  up to 3 different deflection angles should be measurable. 
Note, that while an increase in temperature depopulates  
the Pd$_7$ pentagonal bipyramid triplet ($S$$=$1) GS, the 
singlet ($S$$=$0) state of that structure  and the capped
 octahedron quintet ($S$$=$2) state (Fig.~1F)
gain statistical weight.
Consequently, we predict that for Pd$_7$ 
a rise in temperature would lead first to a decrease of the thermally 
averaged magnetic moment per atom    
$\langle \mu\rangle_{N,T}$$=$$\sum_S 2S\mu_BP_{N,T}(S)/N$ due to a 
sharp increase in the population of the singlet state.
 The subsequent 
 increase of $\langle \mu\rangle_{N,T}$ (see Fig.~3b, T$>$200 K)
 results from the higher thermal population 
of the quintet state relative to the GS
triplet (Fig.~3a).\cite{Fe}
Such anomaly does not  occur in the case  of the Pd$^-_7$ cluster
where the  doublet ($S$$=$1/2) and quartet ($S$$=$3/2)  states  
  start to coexist with the sextet ($S$$=$5/2) GS at 
elevated temperatures (Fig.~3a).
For both neutral and anionic Pd$_{13}$, the 
higher-lying isomers play essentially no role for 
$T$$<$800~K resulting in a weak temperature dependence 
of $\langle \mu\rangle_{N,T}$ (Fig. 3b).

In summary, we found that unlike atomic and
bulk Pd,  both neutral and anionic 
Pd$_N$ clusters ($2\leq N\leq 7$ and $N=13$)  
are  magnetic, with  relatively high LMMs.
Underlying this behavior is 
the hybridization of
atomic $s$ and $d$ states when clusters are
formed, that  depletes
local $d$-contribution around each atom
and leads to an  open-shell-like behavior.
 The abundance of close-lying
SPIs for certain clusters 
should be detectable in thermally controlled 
 SG experiments, and we  predict
an ensemble of Pd$_7$ clusters to exhibit 
an (anomalous) increase of the magnetic moment with temperature.
 The remarkable agreement between the calculated and 
measured vDEs of the cluster anions corroborates the predicted
atomic structures. Our results provide the first
quantitative predictions  pertaining to the
emergence of pronounced magnetic
properties of Pd$_N$ and Pd$^-_N$ clusters, unlike
the weak magnetic tendencies inferred indirectly from 
PES data \cite{Gantefoer}; note in particular
the contrast between our results for Pd$_7$ 
(see Fig. 2d and Fig. 3) and the
suggestion of a zero spin GS
for Pd$_7$ given in Ref. \onlinecite{Gantefoer}.
Furthermore,  our study motivates  temperature-dependent
magnetic deflection (SG) measurements and further investigations of
free and supported Pd clusters, including correlations between
their magnetic properties and their catalytic activity~\cite{Heiz}.

This work is supported by the U.S. DOE, the Deutsche
Forschungsgemeinschaft (M.M.), and the Academy of Finland (H.H.). Computations
have been done on IBM SP2 at the Georgia  Tech Center for
Computational Materials Science and on Cray T3E at NERSC.

%
%

\end{multicols}{}

%
%
\begin{figure}
\caption[]{Binding energies $E_B$ (in eV) of structural (a,b,c) 
and spin ($S$$=$0, 1/2, 1, 3/2 etc.)
isomers of Pd$^-_N$ and Pd$_N$ for $N$$=$1$-$7 (A-F) and 13 (G). 
Filled and open symbols represent anions and neutrals, 
respectively with squares for the ground state
structure (a), circles for structure (b) and triangles for (c). 
The anionic $E_B$ is defined as
the energy per atom to separate Pd$^-_N$ into $N$ 
neutral atoms and an electron.
 The $E_B$ values on the left hand 
side of the panels are mapped on separate 
temperature scales~\cite{Temp} ($T$, in K)
for the anions (top values on the right hand axis of
each panel) and the neutrals  (bottom values), 
giving an estimate of the thermal accessibility
of the isomers. Note the large number of isomers for  
Pd$_7$ and Pd$^-_7$ for $T$$<$400~K.
Some of the structural isomers were not stable
for certain spin values (such cases are denoted
by dashed lines).  
The Pd$^-_4$ square (b) in panel C transformed into a 
rhombus (c) for $S$$=$3/2 and the Pd$_4$ rhombus  transformed 
into a square for $S$$=$0 and 1. The Pd$_6$ 
capped trigonal bipyramid (b) transformed to the 
octahedron (a) for $S$$=$0. The Pd$_7$ capped 
octahedron (b) became a pentagonal bipyramid (a)
for $S$$=$3 and the Pd$_{13}$ and Pd$^-_{13}$ cuboctahedron 
(b) transformed to the icosahedron (a)
for $S$$=$4.5 and 5, respectively. 
}
\end{figure}
%
%
%
%
%
\begin{figure}
\caption[]{Size evolution (panels a-d)  
of energetic and magnetic quantities in Pd clusters.  
(a): binding energy per atom, $E_B$ (in eV),
(b): theoretical (filled squares) 
and  experimental (open circles) 
 vertical electron detachment energy (vDE) 
and the calculated  atom dissociation energy
  $E_D$ of the anions (filled circles).
Experimental values are take from ref.~\cite{Lineberger1,Lineberger2} for 
$N$$=$1$-$3 
(the corresponding results shown for Pd$_{3}^-$ in Fig.~2 of ref.~\cite{Gantefoer}
are essentially the same) 
and from~\cite{Gantefoer} 
for $N$$=$4-7 and 13. The cross and the open circle 
at $N$$=$3 correspond to the maximum of 
peak A and the peak of group B-H in Fig.~1 of ref.~\cite{Lineberger2},
respectively
(and correspondingly for the first and second peaks in Fig.~2 of 
ref.~\cite{Gantefoer}).
For Pd$^-_7$, we show the vDE of the GS ($S$$=$5/2) and a thermally 
accessible SPI ($S$$=$3/2). 
The vDE of Pd$^-_1$ was estimated 
according to Janak's theorem~\cite{Janak} 
by the HOMO energy of Pd$^{-1/2}_1$;
(c): calculated vertical ionization potentials;
(d): the magnetic moment per atom of the GS 
anions (filled squares) and of the GS neutrals (open squares).
(e,f):Density of states (DOS, in 1/eV) of singlet (e) and nonet (f)
icosahedral Pd$_{13}$ cluster.
(g): Constant-value images of spin-polarization density
in singlet, quintet, and nonet icosahedral  Pd$_{13}$ 
(left, middle, right, respectively)
cluster. The purple and yellow denote excess of minority and
majority spin, respectively. Note the transition from antiferromagnetic
to ferromagnetic ordering when going from the singlet to the nonet.}
\end{figure}
%
%
%
%
\begin{figure}
\caption[]{
(a): population probabilities of the different 
SPIs (irrespective the structure) of neutral (dashed curves)
and  anionic (solid curves) heptamers;
(b): thermally averaged magnetic moment 
per atom for Pd$_7$, Pd$^-_7$, Pd$_{13}$ and Pd$^-_{13}$.  }
\end{figure}
%
%

%
%


\begin{thebibliography}{200}
\bibitem{Temp} 
A measure of the accessibility of an isomer is the
temperature $T$ corresponding to the difference between the binding 
energies of the GS and the isomer,
i.e.
$T$$=$$2(NE_{B,GS}$$-$$NE_{B,isomer})/(3N$$-$$6)k_B$
 with the number of cluster atoms $N$ 
and the Boltzman constant $k_B$.
\bibitem{Pd2}
Indeed, the lowest measured~\onlinecite{Lineberger2} electron binding
energy peak for Pd$_3^-$ at 1.66~eV
essentially coincides with the vDE
determined from the PES of Pd$^-_2$ \onlinecite{Lineberger1} (see open circle
for $N$$=$2 in Fig. 2b).  However, the measured PES
peak at 1.66 eV does not exhibit a vibrational structure
characteristic to a dimer~\onlinecite{Lineberger2}.
\bibitem{Fe}   
Such an anomalous behavior has been observed in
magnetic deflection experiments for large
Fe clusters (Ref. 1)
  and it has been conjectured that underlying it
is a structural crystal phase
transformation,
in analogy with the expected coexistence at finite temperature
of the D$_{5h}$
and capped O$_h$ structures found by us for the Pd$_7$ cluster.
\end{thebibliography}
\end{document}